# Modeling of Self-Healing Polymer Composites Reinforced with Nanoporous Glass Fibers

**Vladimir Privman**,\*  **Alexander Dementsov**,  and  **Igor Sokolov**

Center for Advanced Materials Processing and
Department of Physics, Clarkson University,
Potsdam, New York 13699, USA

**Abstract**   We report on our progress towards continuum rate equation modeling, as well as numerical simulations, of self-healing of fatigue in composites reinforced with glue carrying nanoporous fibers. We conclude that with the proper choice of the material parameters, effects of fatigue can be partially overcome and degradation of mechanical properties can be delayed.

\*Web address:  www.clarkson.edu/Privman

In recent years there has been increasing interest in designing "smart materials," specifically, self-healing composites[1-9] that can restore their mechanical properties with time or at least reduce deterioration caused by development of microcracks. One of the popular approaches[1,7-9] has been to use microcapsules broken by developing microcracks and releasing "glue" material into the damaged matrix. For instance, in recent experiments,[1] an epoxy (polymer) was studied, with embedded microcapsules containing a healing agent. Application of a periodic load on a specimen with a crack induced rapture of microcapsules. The healing glue was released from the damaged microcapsules, permeated the crack, and a catalyst triggered a chemical reaction which re-polymerized the crack. We note, however, that a significant amount of microcapsules embedded in the material may actually weaken its mechanical properties and reduce its usable lifetime (though it was noted[1] that small amount of microcapsules actually increased the material toughness). Thus, the density of the "healing" microcapsules is one of the important system parameters to optimize in any modeling approach.

Defects of nanoscale sizes are randomly distributed over the material. Mechanical loads during the use of the material cause formation of craze fibrils along which microcracks develop. This ultimately leads to degradation of the material. Triggering self-healing mechanism *at the nanoscale* offers several advantages for a more effective prevention of further propagation and growth of microcracks, specifically, it is hoped that nanoporous fibers with glue will heal smaller damage features, thus delaying the material *fatigue*. This is in contrast with the earlier studied several-micron size capsules,[1,9] that basically re-glue larger cracks. Furthermore, on the nanoscale the glue is used more efficiently because mixing with the catalyst will be effectively accomplished by diffusion, thereby also eliminating the need for external UV irradiation,[9] etc.

Let us mention our experimental effort, focused on synthesis[6] of *nanoporous* glass capsules for the healing glue. Micron-size glass (silica) capsules consist of arrays of silica nanotubes bundled together into hexagonal fibers. Figure 1 illustrates such syntheses of fibers of 2 μm in diameter and 5 μm in length, with pores of uniform diameters of 3 nm (tunable from 2 nm to 10 nm). Our fibers have mechanical properties suitable for self-healing: They can be easily dispersed in the epoxy matrix, and we have found evidence that they break every time a crack propagates though them.

Presently, *theoretical modeling* and utilization of computational design tools to predict properties of self-healing materials are only in the initiation stages.[10-11] Many theoretical works and numerical simulations,[12-15] consider formation and propagation of large cracks causing irreversible damage to the material. Once developed, a macroscopic crack can hardly be healed by an embedded healing substance. Therefore, the focus of our modeling program has been to consider the time dependence of a gradual formation of damage (fatigue) and its manifestation in material composition, as well as its healing by nanoporous fiber rapture and release of glue.

In this work, we for the first time formulate rate equations for such a process. In addition to continuum rate equations for the material composition, numerical modeling can yield useful information on the structure, and we report first results of Monte Carlo simulations. We also point out that the calculated material composition and structure must be related to macroscopic properties that are experimentally probed. The latter part of the program is relatively well established in Mechanical Engineering,[16] and will not be addressed here. Experimental tests of the fatigue in self-healing epoxies are underway, and results, to be reported elsewhere, are outside the scope of this article.

Let us describe our model of the material composition in the continuum rate equation approach. We denote by $u(t)$ the fraction of material that is undamaged, by $g(t)$ the fraction of material consisting of glue-carrying capsules, by $d(t)$ the fraction of material that is damaged, and by $b(t)$ the fraction of material with broken capsules, so that we have

$$u(t) + g(t) + d(t) + b(t) = 1. \tag{1}$$

We consider the regime of small degree of degradation of the material, i.e., $u(t) \approx 1$, whereas $d(t)$, $b(t)$ and $g(t)$ are relatively small. In fact, $b(0) = 0$.

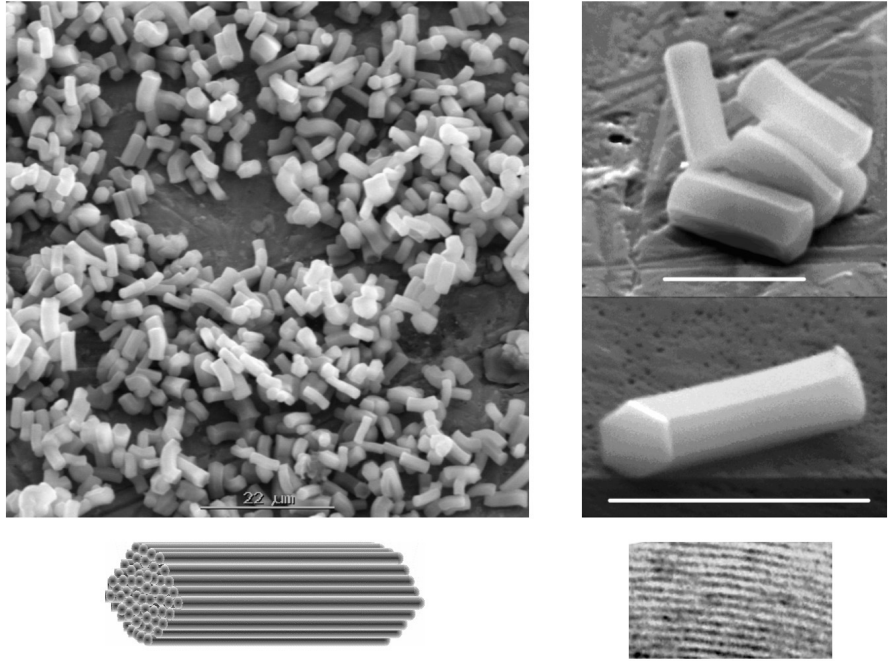

**Fig. 1:** *Experiment*: The top left panel shows a large-area SEM image (bar size 22 μm) of nanoporous glass fibers consisting of self-assembled nanotubes. The top right panel shows zoomed images of fibers (bar size 5 μm). The bottom left drawing is a schematic of the nanotube arrangement, whereas the bottom right panel shows a TEM image near a fiber edge, demonstrating the periodicity of about 3 nm.

For the purposes of simple modeling, we assume that *on average* the capsules degrade with the rate $P$, which is somewhat faster than the rate of degradation of the material itself due to its continuing use (fatigue), $p$, i.e., $P > p$. Thus, we approximately take

$$\dot{g}(t) = -Pg(t), \quad \text{yielding} \quad g(t) = g(0)e^{-Pt}. \tag{2}$$

One can write a more complicated rate equation for $g(t)$, but the added, nonlinear terms are small in the considered regime. However, for the fraction of the undamaged material, we cannot ignore the second, nonlinear term in the relation

$$\dot{u}(t) = -pu(t) + H(t). \tag{3}$$

Here we introduced the healing efficiency, $H(t)$, which can be approximated by the expression

$$H(t) \propto d(t)g(t) \times (\text{volume healed by one capsule}). \tag{4}$$

The healing efficiency is proportional to the fraction of glue capsules, as well as to the fraction of the damaged material, because that is where the healing process is effective. The latter will be approximated by $d(t) \approx 1 - u(t)$, which allows us to obtain a closed equation for $u(t)$. Indeed, in Eq. (3) we can now use

$$H(t) = Ae^{-Pt}[1 - u(t)]. \tag{5}$$

Here the healing efficiency is controlled by the parameter

$$A \propto g(0) \times (\text{volume healed by one capsule}) . \qquad (6)$$

While the model just formulated is quite simple, and many improvement can be suggested, it has the advantage of offering an exact solution,

$$u(t) = u(0) e^{-pt - AP^{-1}(1-e^{-Pt})} + A e^{-pt + AP^{-1} e^{-Pt}} \int_0^t d\tau\, e^{-(P-p)\tau - AP^{-1} e^{-P\tau}} . \qquad (7)$$

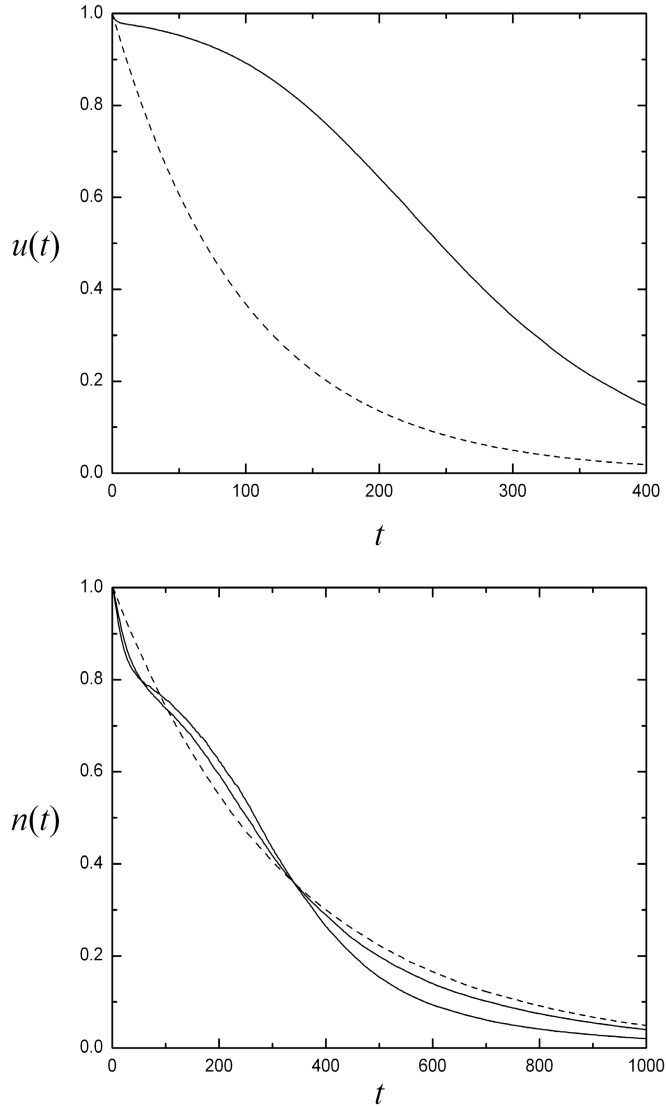

**Fig. 2:** *Top panel*: Example of a calculation of the fraction of the undamaged material according to the rate equation model result Eq. (7). The parameters for the solid curve are $A = 0.5$, $u(0) = 1$, $p = 0.01$, $P=0.02$. The dashed curve corresponds to $A = 0$ (no healing capsules). *Bottom panel*: Fraction of the undamaged cell walls (bonds) obtained in a two-dimensional lattice simulation with short-range healing. The time variable represents the number on Monte Carlo sweeps through the lattice. The bond breakage rates were $p = 0.003$, $P=0.008$. The dashed line corresponds to no healing cells. The solid lines show results with the initial fraction of the healing cells 0.3 and 0.6, with the latter value corresponding to the curve with larger healing for intermediate times, at the expense of more damage for the shortest (and also for large) times.

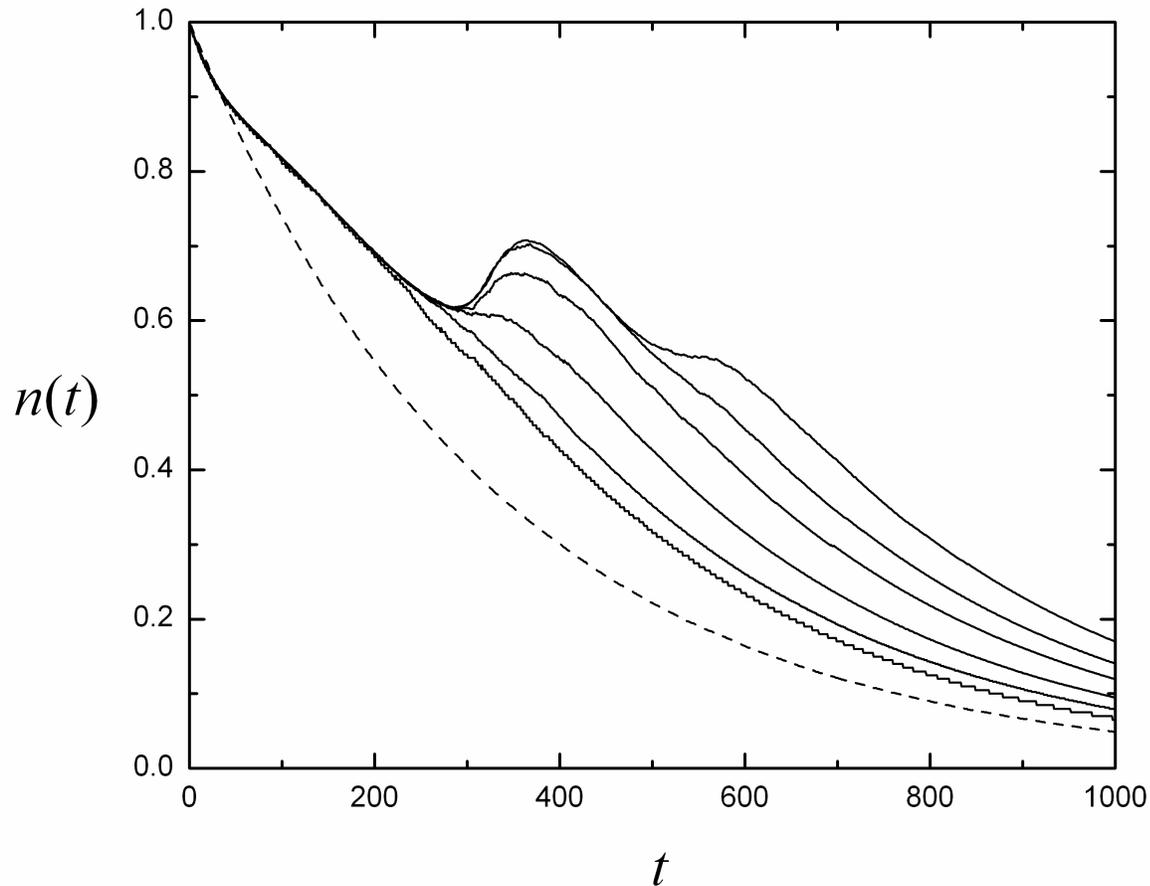

**Fig. 3:** Fraction of undamaged bonds for nonlocal self-healing. Nonmonotonic healing effect and strong dependence on the lattice size are illustrated. Here we took $p = 0.003$ and $P = 0.008$. The dashed curve corresponds to no healing cells. The solid curves are for the initial fraction of the healing cells 0.1, and represent results for several lattice sizes: The healing increases with the lattice size, with the lowest solid curve shown corresponding to size 10×10, then 20×20, 40×40, 80×80, 160×160, and 320×320 (the topmost curve).

An example of this function is shown in the top panel of Fig. 2. The main feature observed is that when the healing efficiency parameter $A$ is not too small, the decay of the fraction of the undamaged material is delayed for some interval of time. This represents the self-healing effect persisting until the glue capsules are used up.

Equation (6) suggests that an important challenge in the design of self-healing materials will be to have the healing effect of most capsules cover volumes much larger than a capsule, in order to compensate for a relatively small value of $g(0)$, which is the fraction of the material volume initially occupied by the glue-filled capsules. Since the glue cannot "decompress," its healing action, after it spreads out and solidifies, should have a relatively long-range stress-relieving effect in order to prevent further crack growth over a large volume. The present simple continuum modeling cannot address the details of the morphological material properties and glue transport; large-scale numerical simulations will be needed to explore this issue.

As a first step towards setting up such a numerical modeling framework, we carried out numerical Monte Carlo simulations on square lattices of varying sizes, with periodic boundary conditions. All the bonds in the lattice are initially present, and the healing cells are distributed uniformly over the lattice. At times $t > 0$, bonds are randomly broken with the probability $p$ for ordinary bonds, and $P > p$ for bonds of healing cells (those with "glue"). If at least two bonds are broken in a healing cell, the glue leaks out and can restore broken bonds. Self-healing was modeled for two limiting cases: The *local* healing, with the glue only spreading to the neighboring cells before solidifying, thus restoring all the bonds of these cells, and the *nonlocal* healing, with the glue propagating as far as it can go via paths that cross broken bonds, to all the cells thus reached from the original healing cell. For each choice of the system parameters, numerical Monte Carlo simulation results were averaged over at least several thousand independent runs.

For *local healing*, a typical result of a numerical simulation for a fraction of undamaged bonds, $n(t)$, is shown in the bottom panel of Fig. 2. There was no significant lattice size dependence noticed for local healing. As the initial density of the healing cells increases, the healing effect for all but the shortest times becomes more profound. However, there is a trade-off in that at very short times the material with more healing cells degrades faster. Indeed, the choice of $P > p$ corresponds to those experimental situations, noted earlier, when the healing cells are easier to break than the surrounding matrix.

For *nonlocal healing*, as the bonds of the lattice break, the effective healing range of the remaining healing cells can actually increase as other healing cells are used up. This occurs because the network of the broken bonds can develop larger clusters before being healed locally by other healing cells. At some times the healing effect can be profoundly magnified, and we also expect strong size-dependence. These features are illustrated in Fig. 3. A nonmonotonic healing effect is observed, which depends on the lattice size.

In summary, we described our first continuum model development and numerical results for self-healing processes, with emphasis on avoiding the initial material degradation (fatigue), as expected for self-healing effected by nanoporous glue-releasing capsules. Our results pose interesting new challenges, primarily related to theoretically understanding the effects of, and the need to increase experimentally, the *range* over which each capsule heals further damage formation in the surrounding matrix once its glue is released, and we hope that we set up the stage for future studies. We gratefully acknowledge support of this research by the US-ARO under grant W911NF-05-1-0339.

# References

1. S. R. White, N. R. Sottos, P. H. Geubelle, J. S. Moore, M. R. Kessler, S. R. Sriram, E. N. Brown, and S. Viswanathan, Nature **409**, 794 (2001).
2. C. Dry, Composite Structures **35**, 263 (1996).
3. B. Lawn, *Fracture of Brittle Solids* (Cambridge University Press, Cambridge, 1993), Chapter 7.
4. C. M. Dry and N. R. Sottos, Proc. SPIE **1916**, 438 (1996).
5. E. N. Brown, N. R. Sottos, and S. R. White, Experimental Mechanics **42**, 372 (2002).
6. Y. Kievsky and I. Sokolov, IEEE Transactions on Nanotechnology **4**, 490 (2005).
7. E. N. Brown, S. R. White, and N. R. Sottos, Journal of Materials Science **39**, 1703 (2004).
8. M. Zako and N. Takano, Journal of Intelligent Material Systems and Structures **10**, 836 (1999).
9. J. W. C. Pang and I. P. Bond, Composites Science and Technology **65**, 1791 (2005).
10. S. R. White, P. H. Geubelle, and N. R. Sottos, *Multiscale Modeling and Experiments for Design of Self-Healing Structural Composite Materials*, unpublished US Air Force research report AFRL-SR-AR-TR-06-0055 (2006).
11. J. Y. Lee, G. A. Buxton, and A. C. Balazs, Journal of Chemical Physics **121**, 5531 (2004).
12. S. Hao, W. K. Liu, P. A. Klein, and A. J. Rosakis, International Journal of Solids and Structures **41**, 1773 (2004).
13. H. J. Herrmann, A. Hansen, and S. Roux, Physical Review B **39**, 637 (1989).
14. M. Sahimi and S. Arbabi, Physical Review B **47**, 713 (1993).
15. J. Rottler, S. Barsky, and M. O. Robbins, Physical Review Letters **89**, 148304 (2002).
16. G. W. Milton, *The Theory of Composites* (Cambridge University Press, Cambridge, 2001), Chapter 10.